\documentstyle[preprint,aps,floats,epsf]{revtex}
\tightenlines
\def\Journal#1#2#3#4{{#1} {\bf #2}, #3 (#4)}
\def\EPJC{Euro. Phys. J. C}
\def\NP{Nucl. Phys.}
\def\PLB{Phys. Lett. B}
\def\PRC{Phys. Rev. C}
\def\PRD{Phys. Rev. D}
\def\PRL{Phys. Rev. Lett.}

\def\ZPA{Z. Phys. A}

\begin{document}
\title{Hadronic scattering of charmed mesons}
\bigskip
\author{Ziwei Lin, C. M. Ko, and Bin Zhang}
\address{Cyclotron Institute and Physics Department, Texas A\&M University,
College Station, Texas 77843}
\maketitle

\begin{abstract}
The scattering cross sections of charm mesons with hadrons such as
the pion, rho meson, and nucleon are studied in an effective
Lagrangian. In heavy ion collisions, rescattering of produced
charm mesons by hadrons affects the invariant mass spectra of 
both charm meson pairs and dileptons resulting from their decays. 
These effects are estimated for heavy ion collisions at
SPS energies and are found to be significant.

\medskip
\noindent PACS number(s): 25.75.-q, 13.75.Lb, 14.40.Lb
\end{abstract}

\section{Introduction}

Recently, experiments on heavy ion collisions at CERN SPS by the
HELIOS-3 \cite{helios3} and NA50 \cite{na50} Collaborations have
shown an enhanced production of dileptons of intermediate masses
($1.5<M<2.5$ GeV). In one explanation, this enhancement is
attributed to dilepton production from secondary meson-meson
interactions \cite{gale}, while in another it was proposed that
dileptons from charm meson decays could also contribute
appreciably in this invariant mass region \cite{imr}. In the
latter case, one of the present authors has shown, based on a
schematic model, that if one assumes that the transverse mass
spectra of charm mesons become hardened as a result of
final state rescattering with hadrons, the invariant mass spectrum of
dimuons from decays of charm mesons would also become hardened, 
and more dimuons would then have an invariant mass between $1.5$ and $2.5$
GeV. Although charm quark production from hadronic interactions
has been extensively studied using perturbative QCD
\cite{combridge,nlo}, not much has been done in studying charm
meson interactions with hadrons. Knowledge on charm meson
interactions with hadrons is important as whether charm mesons
develop a transverse flow depends on how strongly they interact
with other hadrons as they propagate through the matter.

In this paper, we shall first introduce in Sec.~II an effective
Lagrangian to describe the interactions of charm mesons with pion,
rho, and nucleon. Using the coupling constants and cutoff
parameters at the vertices determined either empirically or from
symmetry arguments, we evaluate the scattering cross section of
charm mesons with hadrons. Effects of hadronic scattering on the
charm meson transverse momentum spectrum and the dimuon invariant
mass spectrum from charm meson decays are then estimated in
Sec.~III based on a schematic model for the time evolution of
heavy ion collision dynamics. In Sec.~IV, we summarize our
results and discuss the uncertainties involved in the studies.

\section{Charm meson interactions with hadrons}

\subsection{Lagrangian}

We consider the scattering of charm mesons ($D^+$, $D^-$, $D^0$,
$\bar D^0$, $D^{*+}$, $D^{*-}$, $D^{*0}$, and $\bar D^{*0}$) with
pion, rho, and nucleon. If SU(4) symmetry were exact,
interactions between pseudoscalar and vector mesons could be
described by the Lagrangian  
\begin{eqnarray} 
{\cal L}_{PPV}=ig Tr \left (
P^\dagger V^{\mu \dagger} \partial_\mu P \right ) + {\rm H.c.} \;
\label{lsu4} 
\end{eqnarray} 
where $P$ and $V$ represent, respectively, the
$4\times4$ pseudoscalar and vector meson matrices 
\begin{eqnarray} P&=&
\frac{1}{\sqrt 2}\left (
\begin{array}{cccc}
\frac{\pi^0}{\sqrt 2}+\frac{\eta}{\sqrt 6}+\frac{\eta_c}{\sqrt{12}}
& \pi^+ & K^+ & \bar {D^0} \\
\pi^- & -\frac{\pi^0}{\sqrt 2}+\frac{\eta}{\sqrt 6}+\frac{\eta_c}{\sqrt{12}}
& K^0 & D^- \\
K^- & \bar {K^0} & -\sqrt {\frac{2}{3}}\eta+\frac{\eta_c}{\sqrt{12}}
& D_s^- \\
D^0 & D^+ & D_s^+ & -\frac{3\eta_c}{\sqrt{12}}
\end{array}
\right ) \;, \nonumber \\[2ex]
V&=&\frac{1}{\sqrt 2}\left (
\begin{array}{cccc}
\frac{\rho^0}{\sqrt 2}+\frac{\omega}{\sqrt 6}+\frac{J/\psi}{\sqrt {12}}
& \rho^+ & K^{*+} & \bar {D^{*0}} \\
\rho^- & -\frac{\rho^0}{\sqrt 2}+\frac{\omega}{\sqrt 6}+\frac{J/\psi}
{\sqrt {12}} & K^{*0} & D^{*-} \\
K^{*-} & \bar {K^{*0}} & -\sqrt {\frac{2}{3}}\omega+\frac{J/\psi}{\sqrt {12}}
& D_s^{*-} \\
D^{*0} & D^{*+} & D_s^{*+} & -\frac{3J/\psi}{\sqrt {12}}
\end{array}
\right ) \;. \nonumber
\end{eqnarray}

The above interaction Lagrangian may be considered as being
motivated by the hidden gauge theory, in which there are no
four-point vertices that involve two pseudoscalar mesons and two
vector mesons. This is in contrast to the approach of using the
minimal substitution to introduce vector mesons as gauge
particles, where such four-point vertices do appear. It is,
however, known that the two methods are consistent if one also
includes in the latter approach the axial vector mesons, which are
unfortunately not known for charm hadrons. Furthermore, 
gauge invariance in the latter approach cannot be consistently maintained 
if one uses the experimental vector meson masses, empirical meson 
coupling constants and form factors at interacting vertices.
Expanding the Lagrangian in Eq.~(\ref{lsu4}) in terms of the meson
fields explicitly, we obtain the following Lagrangians for
meson-meson interactions:
\begin{eqnarray} &&{\cal L}_{\pi DD^*}=ig_{\pi DD^*} 
D^{* \mu} \vec \tau \cdot \left [ \bar D ( \partial_\mu \vec \pi) - (
\partial_\mu \bar D) \vec \pi \right ] + {\rm H.c.}\; , \nonumber \\ 
&&{\cal L}_{\rho DD}=ig_{\rho DD} \left [ D \vec \tau (\partial_\mu \bar D)
-(\partial_\mu D) \vec \tau \bar D \right ] \cdot
\vec \rho^\mu \; ,\nonumber \\ 
&&{\cal L}_{\rho \pi\pi}=g_{\rho \pi \pi} \vec
\rho^\mu \cdot \left ( \vec \pi \times \partial_\mu \vec \pi
\right ) \; ,
\end{eqnarray}
where the coupling constants $g_{\pi D D^*}$,
$g_{\rho DD}$, and $g_{\rho\pi\pi}$ are related to the coupling
constant $g$ via the SU(4) symmetry as shown below in
Eq.~(\ref{ppv}).

In SU(3), the Lagrangian for meson-baryon interactions can be
similarly written using the meson and baryon matrices. The
formulation becomes, however, more complicated in SU(4) where a
more general tensor method is required \cite{su4b}. The
interaction Lagrangians needed for our study then include the
following: 
\begin{eqnarray} &&{\cal L}_{\pi NN}=-ig_{\pi NN} \bar N \gamma_5 \vec
\tau N \cdot \vec \pi\; , \nonumber \\ &&{\cal L}_{DN
\Lambda_c}=ig_{DN \Lambda_c} \left (\bar N\gamma_5 \Lambda_c \bar D + \bar
\Lambda_c \gamma_5 N D \right ) \; .  \nonumber
\end{eqnarray}
In the above we have used the following conventions:
\begin{eqnarray}
\pi^\pm=\frac{\pi_1 \mp i \pi_2}{\sqrt 2}, N=(p,n), 
{\rm \;and\;} D=(D^0, D^+). \nonumber 
\end{eqnarray}
Again, SU(4) symmetry would relate the above coupling constants to each
other with the introduction of one more parameter, as shown below in
Eq.~(\ref{mnn}), because there are two SU(4)-invariant
Lagrangians for pseudoscalar meson and baryon interactions.
We also need the following phenomenological Lagrangian:
\begin{eqnarray}
&&{\cal L}_{\rho NN}=g_{\rho NN} \bar N \left ( \gamma^\mu \vec \tau
\cdot \vec \rho_\mu + \frac {\kappa_\rho}{2m_N} \sigma^{\mu \nu} \vec \tau
\cdot \partial_\mu \vec \rho_\nu \right ) N \; , \nonumber
\end{eqnarray}
where values of the coupling constants $g_{\rho NN}$ and $\kappa_\rho$ are
well-known as discussed below.

\subsection{Cross sections}

In Fig.~\ref{diagrams}, Feynman diagrams are shown for charm meson
interactions with the pion (diagrams 1 to 8), the rho meson
(diagrams 9 and 10), and the nucleon (diagrams 11 to 13). Explicit
isospin states are not indicated. The spin and isospin-averaged
differential cross sections for the $t$ channel and $u$ channel
processes can be straightforwardly evaluated, and they are given
by 
\begin{eqnarray} 
\frac{d\sigma_1}{dt} &=& \frac{g_{\rho \pi \pi}^2 g_{\pi DD^*}^2}{32 \pi s
p_i^2 } \frac {\left [
m_\rho^2-2m_\pi^2-2t+(t-m_\pi^2)^2/m_\rho^2 \right ] \left
[ m_{D^*}^2-2m_D^2-2t+(t-m_D^2)^2/m_{D^*}^2 \right ]}
{(t-m_\pi^2)^2} \; , \nonumber \\ \frac{d\sigma_2}{dt} &=&
\frac{1}{3}\frac{d\sigma_1}{dt} \; , \nonumber \\
\frac{d\sigma_3}{dt} 
&=& \frac{g_{\rho \pi \pi}^2 g_{\rho DD}^2}{32 \pi s p_i^2 }
\frac {(2s+t-2m_\pi^2-2m_D^2)^2}{(t-m_\rho^2)^2} \; , \nonumber \\
\frac{d\sigma_{9}}{dt} &=& \frac{1}{3} \frac{d\sigma_1}{dt} \;
,\nonumber \\ \frac{d\sigma_{10}}{dt} &=& \frac{1}{9}
\frac{d\sigma_1}{dt}\; ,\nonumber \\ \frac{d\sigma_{11}}{dt} &=&
\frac{3g_{\pi DD^*}^2 g_{\pi NN}^2}{64 \pi s p_i^2 } \frac {(-t)\left[
m_{D^*}^2-2m_D^2-2t+(t-m_D^2)^2/m_{D^*}^2 \right ]}
{(t-m_\pi^2)^2} \; ,\nonumber \\ \frac{d\sigma_{12}}{dt} &=&
\frac{1}{3}\frac{d\sigma_{11}}{dt}  \; ,\nonumber \\
\frac{d\sigma_{13}}{dt} &=& \frac{3g_{\rho DD}^2 g_{\rho NN}^2}{32 \pi s p_i^2
} \frac {2(1+\kappa_\rho)^2 \left [-su+m_N^2(s+u)+m_D^4-m_N^4 \right ]
-(s-u)^2 \kappa_\rho \left (1+\kappa_\rho/2+t \kappa_\rho/8 m_N^2
\right ) } {(t-m_\rho^2)^2} , \label{cs}
\end{eqnarray}
where $p_i$
denotes the initial momentum of the two scattering particles in
their center-of-mass frame.

For $s$ channel processes through charm meson resonances, shown by
diagrams 4 to 8, the cross section is taken to have a Breit-Wigner
form
\begin{eqnarray} \sigma &=& \frac{(2J+1)}{(2s_1+1)(2s_2+1)}\frac
{4\pi}{p_i^2} \frac{\Gamma_{\rm tot}^2 B_{\rm in}B_{\rm out}}
{(s-M_R^2)^2/s+\Gamma_{\rm tot}^2}\; , \nonumber
\end{eqnarray}
where
$\Gamma_{\rm tot}$ is the total width of the resonance, $B_{\rm
in}$ and $B_{\rm out}$ are their decay branching ratios to the
initial and final states, respectively. We note that diagrams 4 to
7 correspond to processes through the $D_2^*$ and $D_1$ resonances
$D\pi\rightarrow D\pi$, $D\pi\rightarrow D^*\pi$,
$D^*\pi\rightarrow D\pi$ and $D^*\pi\rightarrow D^*\pi$,
respectively, while diagram 8 represents the process $D\pi
\rightarrow D\pi$ through the $D^*$ resonance. Total widths for
$D_2^*$ and $D_1$ resonances are known, and they are
$\Gamma_{D_2^{*0}}=23$ MeV, $\Gamma_{D_2^{*+}}=25$ MeV,
$\Gamma_{D_1^0}=18.9$ MeV, and $\Gamma_{D_1^+}=28$ MeV
\cite{pdbook}. For the width of $D^*$, only an upper limit is
known, i.e., $\Gamma_{D^{*0}}< 2.1$ MeV and $\Gamma_{D^{*+}}<
0.131$ MeV. Studies based on the relativistic potential model
\cite{dwidth} suggest that $\Gamma_{D^{*0}}\simeq 42$ KeV and
$\Gamma_{D^{*+}}\simeq 46$ KeV, and we use these values in this
paper. The branching ratios (BR) are known for $D^{*+}$ and
$D^{*0}$ \cite{pdbook}, but not for $D_2^*$ and $D_1$.
Experimental data show that for both ${D_2^{*0}}$ and ${D_2^{*+}}$
decays one has $\Gamma(D\pi_{ch})/\Gamma(D^*\pi_{ch}) \sim 2$.
Since $D_1$ decays to $D^*\pi$ instead to $D\pi$ due to parity
conservation, we assume BR$(D_1\rightarrow D^*\pi)=1$,
BR$(D_2^*\rightarrow D^*\pi)=1/3$, and BR$(D_2^*\rightarrow
D\pi)=2/3$, neglecting possible decays of $D_1$ and $D_2^*$ to
$D\rho$ and $D^*\rho$, respectively \cite{brd}.

\subsection{Coupling constants}

For coupling constants, we use the empirical values $g_{\rho \pi \pi}=6.1$
\cite{rpp}, $g_{\pi DD^*}=4.4$, $g_{\rho DD}=2.8$\cite{rdd}, $g_{\pi NN}=13.5$
\cite{pnn}, $g_{\rho NN}=3.25$, and $\kappa_\rho=6.1$ \cite{rnn}.
From SU(4) symmetry, as assumed in the Lagrangian in
Eq.~(\ref{lsu4}), one would expect the following relations among
these couplings constants: 
\begin{eqnarray} g_{\pi KK^*} (3.3)=g_{\pi DD^*}
(4.4)=g_{\rho KK} (3.0)=g_{\rho DD} (2.8) 
=\frac {g_{\rho \pi \pi}}{2} (3.0)\; .
\label{ppv}
\end{eqnarray}
One sees that the empirical values given in
parentheses agree reasonably well with the prediction from SU(4)
symmetry. Signs of the coupling constants are not specified as the
possible interferences among diagrams 3, 4, and 8 are not included.
We note that the coupling constant $g_{\pi DD^*}$ is consistent with that
determined from the $D^*$ width 
\begin{eqnarray} 
\Gamma_{D^*\rightarrow \pi
D}=\frac {g_{\pi DD^*}^2 p_f^3}{2\pi m_{D^*}^2} \; , \nonumber 
\end{eqnarray} 
where $p_f$ is the momentum of final particles in the $D^*$ rest frame.

\subsection{Form factors}

To take into account the structure of hadrons, we introduce form
factors at the vertices. For $t$ channel vertices, monopole form
factors are used, i.e., 
\begin{eqnarray} f(t)&=&\frac
{\Lambda^2-m_\alpha^2}{\Lambda^2-t}\;, \nonumber
\end{eqnarray}
where
$\Lambda$ is a cutoff parameter, and $m_\alpha$ is the mass of
the exchanged meson. For cutoff parameters, we use the empirical
values $\Lambda_{\rho \pi \pi}=1.6$ GeV \cite{rpp}, $\Lambda_{\pi
NN}=1.3$ GeV, and $\Lambda_{\rho NN}=1.4$ GeV\cite{pnn}. However,
there are no experimental information on $\Lambda_{\pi DD^*}$ and
$\Lambda_{\rho DD}$, and their values are assumed to be similar to
those determined empirically for strange mesons, i.e.,
$\Lambda_{\pi DD^*}=\Lambda_{\pi KK^*}=1.8$ GeV, $\Lambda_{\rho
DD}=\Lambda_{\rho KK}=1.8$ GeV \cite{rpp}. For $s$ channel
processes, shown in diagrams 4 to 8, that are described by
Breit-Wigner formula, no form factors are introduced.

\subsection{On-shell divergence}

The cross sections in Eq.~(\ref{cs}) for diagrams 2 and 9 ($D\rho
\leftrightarrow D^* \pi$) are singular because the exchanged
mesons can be on-shell. Since the on-shell process describes a
two-step process, their contribution needs to be subtracted from
the cross section. This can be achieved by taking into account the
medium effects which add an imaginary self-energy to the mass of
the exchanged pion as in Ref. \cite{singular}.  We take the
imaginary pion self-energy to be $50$ MeV and have checked that
the calculated thermal average of the cross sections do not change
much with values between $5$ and $500$ MeV. We note that there are
other ways to regulate this singularity \cite{baier}.

\subsection{Thermal average}

We are interested in the thermal averaged cross sections for the
processes shown in Fig.~\ref{diagrams}. For a process $1+2
\rightarrow 3+4$, where the initial-state particles $1$ and $2$
are both described by thermal distributions at temperature $T$,
the thermal averaged cross section is given by
\begin{eqnarray} 
\langle \sigma v \rangle
&=&\frac {\int_{z_0}^{\infty} dz \left [z^2-(\alpha_1+\alpha_2)^2
\right ] \left [z^2-(\alpha_1-\alpha_2)^2 \right ] K_1(z) \;
\sigma (s=z^2T^2) } {4(1+\delta_{12}) \alpha_1 K_2(\alpha_1)
\alpha_2 K_2(\alpha_2)} \;. \nonumber
\end{eqnarray} 
In the above, $\alpha_i
=m_i/T$ ($i=1$ to $4$), $z_0={\rm
max}(\alpha_1+\alpha_2,\alpha_3+\alpha_4)$, $\delta_{12}$ is $1$
for identical initial-state particles and $0$ otherwise, and $v$
is their relative velocity in the collinear frame, i.e., 
\begin{eqnarray} v=
\frac {\sqrt {(k_1\cdot k_2)^2-m_1^2 m_2^2}}{E_1 E_2}\;. \nonumber
\end{eqnarray}

In Fig.~\ref{sv}(a), we show the results for the thermal averaged
cross sections as functions of temperature. It is seen that
dominant contributions are from $D$ and $D^*$ scatterings by
nucleon, $D$ scattering by pion via rho exchange, and $D$
scattering by rho meson via pion exchange. In obtaining these
result, the rho meson mass is taken at its peak value of $770$
MeV.

\section{Estimates of rescattering effects}

As shown in the schematic model of Ref. \cite{imr}, if one assumes
that charm mesons interact strongly in the final state hadronic matter, then
their transverse mass ($m_\perp$) spectra 
and pair invariant mass spectra would become harder than
the initial ones as a result of the appreciable transverse flow of
the hadronic matter. Dilepton decays of charm mesons would then
lead to an enhanced yield of intermediate-mass dileptons in heavy
ion collisions. 
In this section, we estimate the effects of hadronic rescattering on
charm meson $m_\perp$ spectra and the invariant-mass distribution
of dileptons from their decays in heavy ion collisions at SPS
energies.

To characterize the scattering effects on charm mesons, we first
determine the squared momentum transfer to a charm meson when it
undergoes a scattering process $D_1 X_1 \rightarrow D_2 X_2$. In
the rest frame of $D_1$, the squared momentum of the final charm
meson $D_2$ is given by 
\begin{eqnarray} p_0^2=\frac {\left
[(m_{D_1}+m_{D_2})^2-t \right ] \left [(m_{D_1}-m_{D_2})^2-t\right
]}{(2 m_{D_1})^2} \nonumber 
\end{eqnarray} 
for $t$ channel processes with
four momentum transfer $t$. For $u$ channel processes, one
replaces $u$ for $t$ in the above expression.

We determine the total number of collisions suffered by a charm
meson from its scattering cross sections and the time evolution of
the hadron densities. In the charm meson local frame, we assume
that the density evolution of hadrons is inversely proportional to
the proper time, i.e.,
\begin{eqnarray} \rho(\tau) \propto \frac{1}{\tau}\;.
\label{drdt} 
\end{eqnarray} 
Neglecting the effect of transverse expansion on
the density evolution, the total number of scatterings for a charm
meson is then 
\begin{eqnarray} N &=& \int_{\tau_0}^{\tau_F} \sigma v \rho
d\tau = \sigma v \rho_0 \tau_0 \ln \left ( \frac {\tau_F}{\tau_0}
\right ) \nonumber \\ &=& \sigma v \rho_0 \tau_0 \ln \left ( \frac
{t_F}{\tau_0 \cosh y} \right ) \simeq \sigma v \rho_0 \tau_0 \ln
\left ( \frac {R_\perp m_\perp}{\tau_0 p_\perp} \right ) \;,
\nonumber
\end{eqnarray}
which leads to the following thermal average of the
squared total momentum transfer due to scatterings,
\begin{eqnarray} 
\langle p_{\rm S}^2 \rangle
=\langle N p_0^2 \rangle= \left[ \sum_{i=\pi,\rho,N\cdots}\langle \sigma v
p_0^2 \rangle_i\rho_{i0}\right ] \tau_0 \ln \left ( \frac {R_\perp
m_\perp}{\tau_0 p_\perp} \right ) \;. \label{total}
\end{eqnarray}
In
obtaining the above result, we have assumed the same initial and
final proper times for the time evolution of different particle
species that are involved in the scattering. Equation~(\ref{total})
shows that the relevant quantity is $\langle \sigma v p_0^2 \rangle$ instead of
the usual $\langle \sigma v \rangle$. We show in Fig.~\ref{sv}(b) this thermal
average for all scattering channels considered in the present
study. It is seen that the dominant contributions to $\langle \sigma v \rangle$
remain important for $\langle \sigma v p_0^2 \rangle$ and the process involving
$D^*$ scattering by rho meson via pion exchange also becomes
significant.

Summing up contributions from the scattering channels in
Fig.~\ref{diagrams}(a), \ref{diagrams}(b), and \ref{diagrams}(c) separately,
and simply dividing by 2 to account for the average over $D$ and $D^*$, we
obtain, at $T=150$ MeV,
\begin{eqnarray} \langle \sigma v p_0^2 \rangle \simeq 1.1, 1.5, {\rm
\;and \;} 2.7 {\rm \;mb~GeV^2} \nonumber
\end{eqnarray}
for $\pi$,
$\rho$, and $N$ scatterings with charm mesons, respectively.

For central Pb+Pb collisions at SPS energies, the initial
particle numbers can be obtained from Ref. \cite{gq}, i.e., there
are 500 $\pi$, 220 $\rho$, 100 $\omega$, 80 $\eta$, 180 $N$, 60
$\Delta$, and 130 higher baryon resonances. The initial densities
at central rapidity can then be estimated using $\rho_0 \tau_0
\simeq (dN/dy)/(\pi R_A^2) \simeq N/(4 \pi R_A^2)$. For a
conservative estimate on the scattering effect, we only include
$\pi$, $\rho$, and nucleon. The initial densities for pion, rho
meson, and nucleon are thus
\begin{eqnarray} \rho_0 \tau_0 \simeq 0.79, 0.35,
{\rm \;and \;} 0.28 {\rm \;fm^{-2}} \; , \nonumber
\end{eqnarray}
respectively. Equation~(\ref{total}) then gives
\begin{eqnarray} 
\langle p_{\rm S}^2 \rangle
&\simeq & \left [ \langle \sigma v p_0^2 \rangle_\pi \rho_{\pi 0} \right .
+\langle \sigma v p_0^2 \rangle_\rho \rho_{\rho 0} \left . 
+\langle \sigma v p_0^2 \rangle_N
\rho_{N 0} \right ] \tau_0 \ln \left ( \frac {R_\perp
\langle m_\perp \rangle}{\tau_0 \langle p_\perp \rangle} \right ) \nonumber\\ 
&\simeq & 
(1.1 \times 0.79 + 1.5\times 0.35 + 2.7 \times 0.28)/10 \times \ln 16.7
\simeq 0.61 \;(\rm GeV^2) \;. \label{min}
\end{eqnarray}
In the above, we
have taken $\tau_0=1$ fm and $R_\perp \simeq R_A \simeq 1.2
A^{1/3}$ fm. 
We have also used the relations 
$\langle p_\perp \rangle \sim \sqrt
{\langle p_\perp^2 \rangle} \simeq \sqrt {2m T_{\rm eff}}$ 
and $\langle m_\perp \rangle \simeq
m + T_{\rm eff}$ as given by Eq.~(\ref{pp2}) in Appendix A. Since
the charm meson $T_{\rm eff}$ increases as a result of the
rescatterings, $\langle p_{\rm S}^2 \rangle$ needs to be determined
self-consistently. However, because of the logarithmic dependence
shown in Eq.~(\ref{min}), 
$\langle p_{\rm S}^2 \rangle$ is not very sensitive to
the value of $T_{\rm eff}$, and we have taken $T_{\rm eff}=200$
MeV in obtaining the above numerical results. We note that even
though pions appear to be less important in Fig.~\ref{sv}, their
contribution to the rescattering effect is important due to their
high densities, as evident from the numerical values shown in
Eq.~(\ref{min}).

The total squared momentum transfer from hadronic scatterings as
given by Eq.~(\ref{total}) can be characterized by a temperature
parameter $T_{\rm S}$, defined by Eq.~(\ref{pth2}) in Appendix A.
Using the values given in Eq.~(\ref{min}), we obtain $T_{\rm S}
\simeq 96$ MeV from Eq.~(\ref{pth2nlo}) of Appendix A.
From Fig.~\ref{enhance}, which relates $T_{\rm S}$ to $T_{\rm eff}$
and to the enhancement factor $R$ for dimuons from charm meson
decays into the NA50 acceptance, this gives an effective inverse
slope parameter of $T_{\rm eff}=235$ MeV for the final charm meson
$m_\perp$ spectrum if the initial one is taken to be $160$ MeV,
and a dimuon enhancement factor of about $2.1$ is obtained.

\section{Summary and discussions}

In summary, we have calculated the cross sections for scatterings
between charm mesons and hadrons such as pion, rho meson, and
nucleon. Hadronic scatterings of charm mesons in heavy ion
collisions can significantly affect the charm meson spectra and
the dilepton spectra from charm meson decays. An estimate of this
effect in heavy ion collisions at SPS energies is given, and we
find that it leads to a hardened charm meson spectra and an
enhanced intermediate-mass dileptons from charm meson decays.
These results thus give a more quantitative justification of the
arguments proposed in Ref. \cite{imr}.

However, the results obtained in the present study are still
incomplete as we have not included diagrams involving the exchange
of heavier particles such as charm hadrons. The scattering cross
sections between charm mesons and hadrons such as kaon, $\omega$,
$\eta$, $\Delta$, and higher baryon resonances are not calculated
either. Furthermore, we have not calculated the contribution due
to diagrams shown in Fig.~\ref{diagrams2}, where charm mesons
scatter with nucleons via a $\Lambda_c$ exchange. There is a large
uncertainty in their contributions as no empirical information on
the coupling constant $g_{DN \Lambda_c}$ is available. The SU(4) symmetry
gives
\begin{eqnarray} 
g_{DN \Lambda_c}=g_{KN\Lambda}=\frac{3-2\alpha_D}{\sqrt 3} g_{\pi NN}
\simeq g_{\pi NN}=13.5 \;, \label{mnn}
\end{eqnarray}
where $\alpha_D=D/(D+F)$
with $D$ and $F$ being the coefficients for the $D$-type and
$F$-type coupling, and $\alpha_D \simeq 0.64$ \cite{alphad}. On
the other hand, QCD sum rule studies suggest a smaller value
$g_{DN \Lambda_c} \simeq 6.7 \pm 2.1$ \cite{sumrule}. Because of this
uncertainty in these two processes involving the $DN\Lambda_c$
coupling, we choose to leave them out in our study. In the future,
when empirical information on this parameter is known from $DN$
scattering, then processes involving the $DN\Lambda_c$ coupling
can be addressed.

We note that the estimates given above are based on a simple
assumption on the time evolution of the dense hadronic system,
which enables us to make an analytical estimate of the
rescattering effects. As a result, we have neglected the
transverse expansion of the hadronic system which would lead to a
faster decrease of hadron densities than the linear dropping
assumption in Eq.~(\ref{drdt}). We have also neglected the
chemical equilibration processes which, e.g., may decrease the
total number of rho mesons and increase that of pions as a
function of time \cite{gq}.

Moreover, we have used only the isospin averaged cross sections
and also averaged the rescattering effects on $D$ and $D^*$
mesons. Without a full cascade calculation and fully treating the
isospin, we do not know the final composition of charm mesons,
e.g, the ratios $D^*/D$ and $D^0/D^+$. A naive expectation gives
$D^*/D=3$, and consequently $D^0/D^+ \simeq 3$\cite{hvq}. However,
even for $pp$ collisions the relative weights of produced charm
mesons are not well measured experimentally. We emphasize that the
charm meson composition could have a sizable effect on the lepton
and dilepton yields from charm decays, because $D^+$ and $D^0$
have very different branching ratios for semileptonic decays
($17.2\%$ from $D^+$ and $6.7\%$ from $D^0$).

In a hadronic cascade model, the time evolution and the chemical
equilibration of the hadronic system can be simulated much better.
Using cross sections with the full isospin information, and
keeping track of the charm meson isospins during scatterings, the
final charm meson composition can be determined. Therefore,
further studies based a cascade code along these directions are
much needed for a quantitative study of the rescattering effects
on charm meson observables.

For heavy ion collisions at RHIC energies, a dense partonic system
is expected to be formed during the early stage of the collision.
In addition to hadronic rescatterings of charm mesons, partonic
rescattering effects on charm quarks also need to be included.
Furthermore, radiative processes of charm quarks inside the QGP
would further complicate the issue as they may cause energy loss
\cite{bdmps} and soften the charm meson $m_\perp$ and pair invariant mass 
spectra \cite{eloss}. Therefore, more studies are needed before one can
make predictions for RHIC.

\section*{Acknowledgments} 
We thank S. Vance, X.-N. Wang and
Pang Yang for helpful discussions. This work was supported in part
by the National Science Foundation under Grant No. PHY-9870038,
the Welch Foundation under Grant No. A-1358, and the Texas
Advanced Research Project No. FY97 010366-068.

\section*{Appendix}
\setcounter{equation}{0}
\def\theequation{A\arabic{equation}}

In this appendix, we derive the relation between the total squared
momentum transfer to charm mesons due to hadronic scatterings and
the increase of the inverse slope of charm meson $m_\perp$
spectra.
Consider a charm meson at central rapidity with an initial
transverse momentum $p_{\perp \rm I}$ along the $y$ axis. After a
scattering which gives the charm meson a momentum $\vec p_{\rm S}$
in its rest frame, its final transverse momentum is given by
\begin{eqnarray}
p_{x\rm F}=p_{x\rm S}\; , \; p_{y\rm F}=\gamma_{\rm I} (p_{y\rm
S}+\beta_{\rm I} E_{\rm S}) \;, \nonumber
\end{eqnarray}
where
\begin{eqnarray}
\beta_{\rm I}=\frac {p_{\perp \rm I}}{\sqrt {p_{\perp \rm
I}^2+m^2}}\; , \gamma_{\rm I}=\frac {1}{\sqrt {1-\beta_{\rm I}^2}}
\;. \nonumber
\end{eqnarray}
Assuming that $\vec p_{\rm S}$ is isotropic in
the charm meson rest frame, the average of the squared final
transverse momentum of the charm meson is then related to that of
the squared initial transverse momentum by
\begin{eqnarray} 
\langle p_{\perp \rm F}^2 \rangle= \langle p_{\perp \rm I}^2 \rangle
+ \left ( \frac{2}{3}+\frac{4\langle p_{\perp \rm I}^2 \rangle}{3m^2}\right ) 
\langle p_{\rm S}^2 \rangle \;, \label{relation}
\end{eqnarray}
where 
$\langle p_{\rm S}^2 \rangle$ is the average of the squared total momentum
transfer to the charm meson as given by Eq.~(\ref{total}). For an
isotropic $\vec p_{\rm S}$ distribution, Eq.~(\ref{relation}) is
actually true for a charm meson at any rapidity.

If we parametrize the $m_\perp$ spectrum of charm mesons as
\begin{eqnarray}
\frac{dN}{m_\perp dm_\perp} \propto e^ {-m_\perp/T_{\rm eff}}
\nonumber
\end{eqnarray}
in terms of an inverse slope parameter $T_{\rm eff}$, then
\begin{eqnarray}
\langle p_\perp^2 \rangle=2 T_{\rm eff}^2
\left ( \frac{m}{T_{\rm eff}}+2+\frac{1}{m/T_{\rm eff}+1}
\right ) \;.
\label{pp2}
\end{eqnarray}

As in the schematic study of Ref. \cite{imr}, we characterize the
scattering strength $\langle p_{\rm S}^2 \rangle$ by an equivalent temperature
parameter $T_{\rm S}$ via
\begin{eqnarray} 
\langle p_{\rm S}^2 \rangle &&=\frac{\int p^2
e^{-E/T_{\rm S}} d^3p}{\int e^{-E/T_{\rm S}} d^3p} =T_{\rm S}^2 \;
\frac{\int x^4 \exp {\left (-\sqrt {x^2+(m/T_{\rm S})^2}\right )}
dx} {\int x^2 \exp {\left (-\sqrt {x^2+(m/T_{\rm S})^2}\right )}
dx} \nonumber \\ &&= 3T_{\rm S}^2 \left [\frac{m}{T_{\rm
S}}+\frac{5}{2}+{\cal O}\left (\frac{1}{m/T_{\rm S}}\right)\right
]. \label{pth2} 
\end{eqnarray}

Both $T_{\rm S}$ and $T_{\rm eff}$ are expected to be small
compared with the charm meson mass ($m \simeq 1.87$ GeV).
Keeping only the leading term in Eqs.~({\ref{pp2}) and
({\ref{pth2}) then gives $\langle p_\perp^2 \rangle=2 m T_{\rm eff}$
and $\langle p_{\rm S}^2 \rangle=3 m T_{\rm S}$ in the nonrelativistic limit.
Equation~(\ref{relation}) thus gives
\begin{eqnarray}
T_{\rm eff}^{\rm F}
&\simeq& T_{\rm eff}^{\rm I} + T_{\rm S} \;.
\label{lo}
\end{eqnarray}
If we also keep the next-to-leading term in Eqs.~({\ref{pp2})
and ({\ref{pth2}), we then obtain
\begin{eqnarray}
\langle p_{\rm S}^2 \rangle \simeq 3m T_{\rm S} + \frac{15}{2} T_{\rm S}^2
\label{pth2nlo}
\end{eqnarray}
and
\begin{eqnarray}
T_{\rm eff}^{\rm F} + \frac{2 {T_{\rm eff}^{\rm F}}^2}{m}
&\simeq& T_{\rm eff}^{\rm I} + \frac{2 {T_{\rm eff}^{\rm I}}^2}{m}
+ T_{\rm S} \left (1+\frac {5T_{\rm S}}{2m} \right )
\left [1+\frac{4T_{\rm eff}^{\rm I}}{m} \left (1+\frac{2T_{\rm eff}^{\rm I}}
{m} \right ) \right ]\;.
\label{nlo}
\end{eqnarray}

For central Pb+Pb collisions at SPS energies, the relations
given by Eqs.~(\ref{lo}) and (\ref{nlo}) are shown in
Fig.~\ref{enhance} together with the results obtained from the
Monte Carlo simulations in the schematic study of Ref. \cite{imr}.
These relations agree qualitatively with that from the
simulations. However, they differ quantitatively, because rapidity
changes due to rescatterings are not taken into account in the
present analytical estimates.

\pagebreak
{}

\pagebreak
\begin{figure}[ht]
\setlength{\epsfxsize=0.75\textwidth}
\setlength{\epsfysize=0.25\textheight}
\centerline{\epsfbox{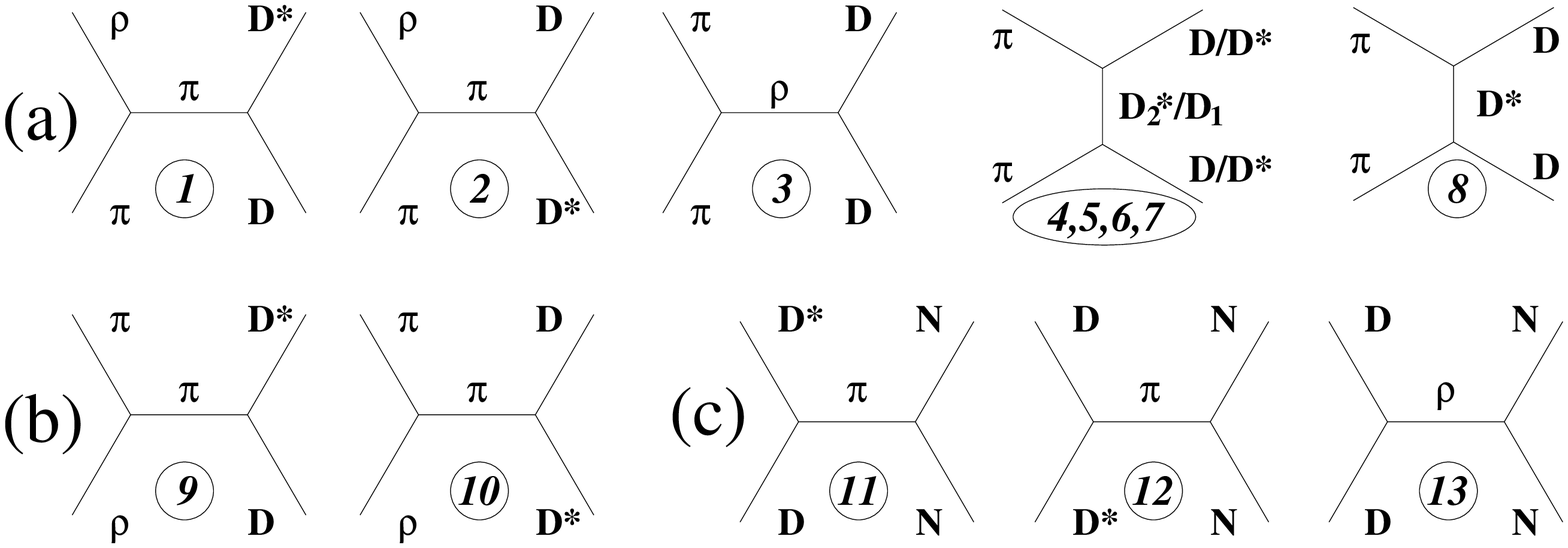}}
\vspace{1cm}
\caption{ Feynman diagrams for (a) $D \pi$,
(b) $D \rho$, and (c) $D N$ scatterings. }
\label{diagrams}
\end{figure}

\pagebreak
\begin{figure}[ht]
\setlength{\epsfxsize=0.75\textwidth}
\setlength{\epsfysize=0.7\textheight}
\centerline{\epsfbox{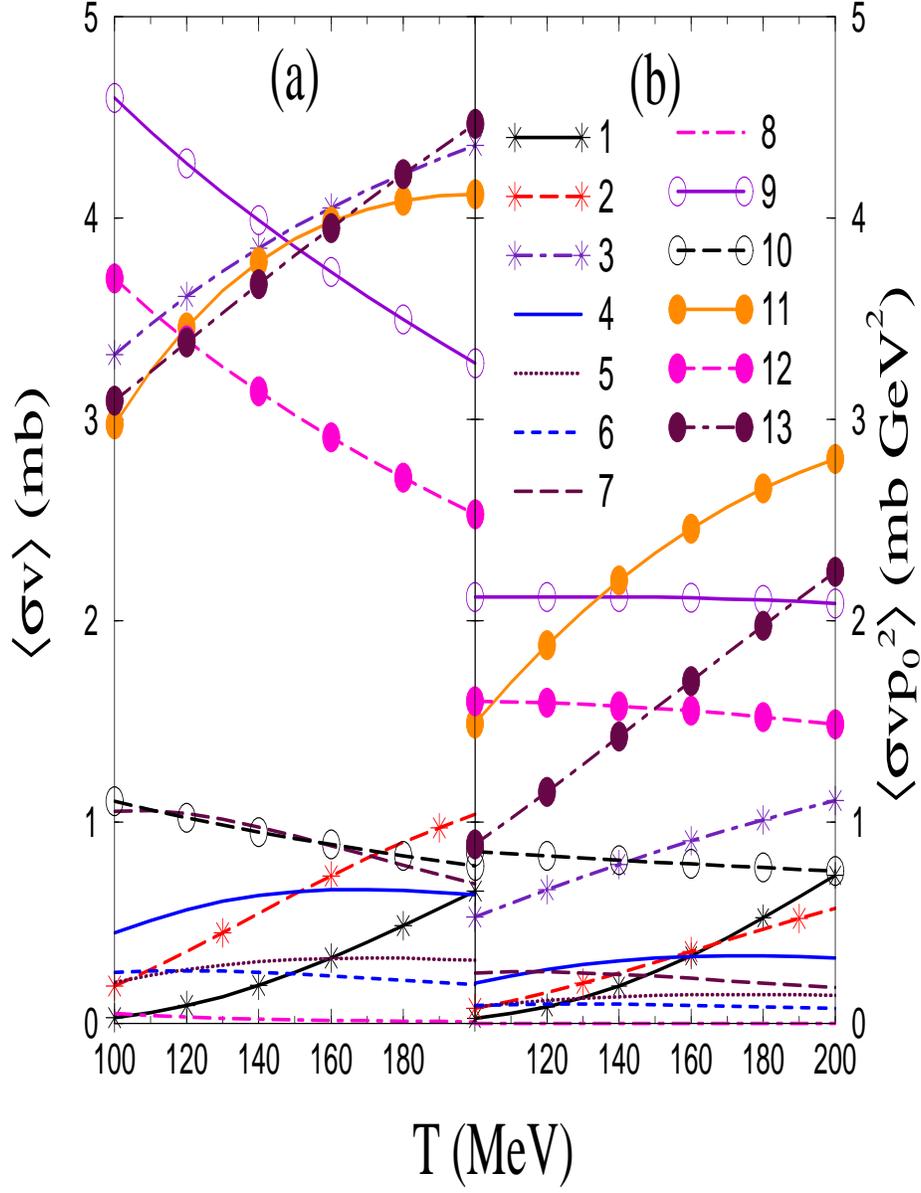}}
\vspace{1cm}
\caption{
Thermal average (a) $\langle \sigma v \rangle$,
and (b) $\langle \sigma v p_0^2 \rangle$
of charm meson scattering cross sections as functions of temperature.
Numbers labeling the curves correspond
to the diagram numbers in Fig.~\ref{diagrams}. }
\label{sv}
\end{figure}

\pagebreak
\begin{figure}[ht]
\setlength{\epsfxsize=0.75\textwidth}
\setlength{\epsfysize=0.7\textheight}
\centerline{\epsfbox{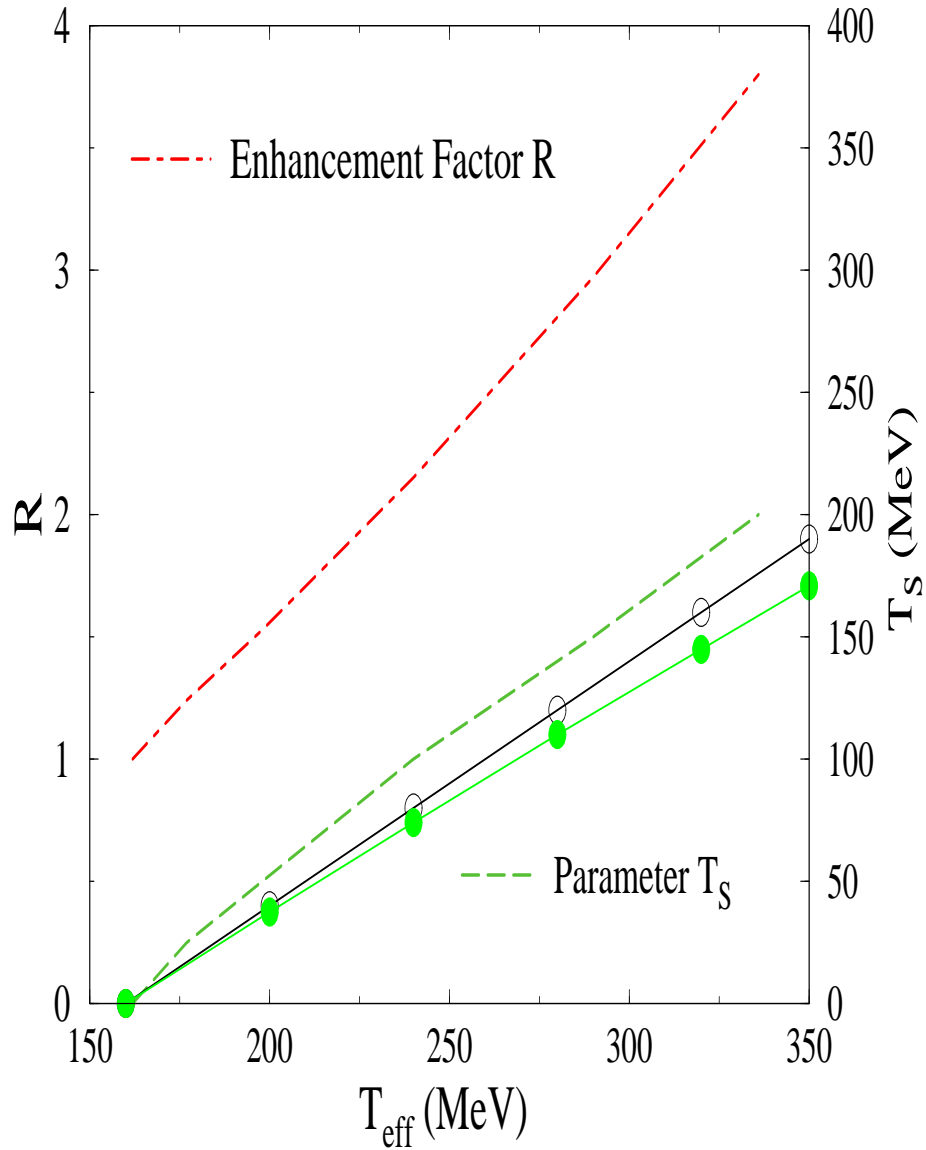}}
\vspace{1cm}
\caption{
The dimuon enhancement factor $R$ within the simulated NA50
acceptance and the equivalent temperature parameter $T_{\rm S}$ due to
scatterings as functions of the final inverse slope of charm mesons
$T_{\rm eff}$ \protect\cite{imr}. The curve with open circles is
from Eq.~(\ref{lo}), while the curve with filled circles is from
Eq.~(\ref{nlo}). }
\label{enhance}
\end{figure}

\pagebreak
\begin{figure}[ht]
\setlength{\epsfxsize=0.75\textwidth}
\setlength{\epsfysize=0.3\textheight}
\centerline{\epsfbox{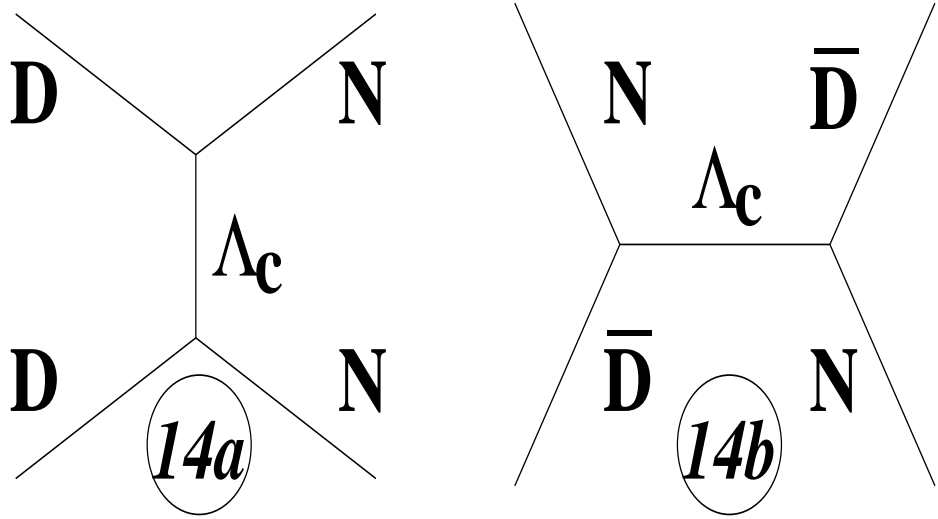}}
\vspace{1cm}
\caption{ Feynman diagrams for charm meson scattering with nucleon via the
$\Lambda_c$ exchange.
}
\label{diagrams2}
\end{figure}

\end{document}